# Generating coherent soft x-ray pulses in the water window with a high-brightness seeded free-electron laser

Kaishang Zhou, Chao Feng*, Haixiao Deng, and Dong Wang

*Shanghai Institute of Applied Physics, Chinese Academy of Sciences, Shanghai 201800, China*

* Corresponding author: fengchao@sinap.ac.cn

**Abstract** We propose a new scheme to generate high-brightness and temporal coherent soft x-ray radiation in a seeded free-electron laser. The proposed scheme is based the coherent harmonic generation (CHG) and superradiant principles. A CHG scheme is first used to generate coherent signal at ultra-high harmonics of the seed. This coherent signal is then amplified by a series of chicane-undulator modules via the fresh bunch and superradiant processes in the following radiator. Using a representative of realistic set of parameters, three-dimensional simulations have been carried out and the simulations results demonstrated that 10 GW-level ultra-short coherent radiation pulses in the water window can be achieved by using the proposed technique.

PACS numbers: 41.60.Cr

## I. INTRODUCTION

Intense, ultra-short, coherent soft x-ray radiation pulses generated by free-electron lasers (FELs) are becoming an essential tool to achieve significant breakthroughs in various scientific domains such as femto-chemistry, material science, biology and so on. Currently, most of the presently existing or planned x-ray FEL facilities [1-6] are based on the self-amplified spontaneous emission (SASE) principle [7, 8]. While the radiation from a SASE FEL always has good spatial coherence, it typically has rather limited temporal coherence and large shot-to-shot fluctuations. There are a number of scientific applications, such as resonant scattering and spectroscopic techniques, that require, or could benefit from, a better temporal coherence.

In order to improve the temporal properties, several SASE-based techniques have been developed in recent years [9-14]. However, the final output pulses from these schemes still suffer from the intrinsic chaotic properties of SASE and always have large intensity or central wavelength fluctuations. An alternative way to significantly improve the temporal coherence of SASE relies on the manipulation of the electron beam longitudinal phase space through external seeding schemes, such as high-gain harmonic generation (HGHG) [15], echo-enabled harmonic generation (EEHG) [16, 17] or phase-merging enhanced harmonic generation [18, 19]. Here after, we will be referring to the external seeding schemes simply as seeded FELs. Seeding the FELs with external coherent laser pulses has the additional advantage of providing radiation pulses with well-defined timing with respect to the seed laser, thus allowing pump-probe experiment to be performed with high temporal resolution. What's more, in a seeded FEL, one can manipulate the longitudinal phase correlation and control the relative phase of the radiation pulses, which is necessary for some phase-related experiments [20, 21].

One drawback of seeded FELs is that the wavelength cover range is limited due to the lack of suitable seeding laser at short wavelength. For a commercial UV laser with central wavelength around 260 nm, the harmonic up-conversion number of a single stage HGHG is about one order of magnitude lower than what is required for reaching the x-ray region. Further extension of the output to shorter wavelength needs a larger energy modulation amplitude, which will degrade the quality of the electron beam and results in a lower FEL output power in the following radiator. When the laser induced energy spread is too large, the FEL will work in the coherent harmonic generation (CHG) regime [15], where the power amplification process will be quickly saturated and no exponential growth is expected in the radiator. In order to improve the frequency multiplication efficiency of a seeded FEL with small laser induced energy spread, more complicated schemes such as cascaded HGHG or EEHG have been developed [22-26].

In this paper, we propose a more straightforward method, termed high-brightness HGHG (HB-HGHG), to generate intense soft x-ray radiation in a seeded FEL. The proposed technique combines the CHG technique with the superradiant method [27-30]. The ultra-short high harmonic coherent signal generated from a CHG is continually amplified by the fresh part of the electron beam in a radiator with a series of temporal shifters. When compared with the previous seeded FEL schemes, the proposed technique has the advantages of simple setup, ultra-high harmonic up-conversion efficiency and high output brightness, simultaneously. This scheme can be easily implemented at already existing or planed seeded FEL facilities, such as FERMI FEL [22] or the Shanghai soft x-ray FEL facility (SXFEL) [31].

This paper is organized as follows: Sec. II shows the layout and principle of the HB-HGHG. Sec. III discusses the requirements for the seed laser and the delay chicanes to realize the proposed technique. With realistic electron beam parameters, optimized start-to-end simulations for a soft x-ray FEL facility based on the HB-HGHG are given in Sec. IV. Finally, the conclusions are summarized in Sec. V.

## II. SCHEMATIC DESCRIPTION OF THE HB-HGHG

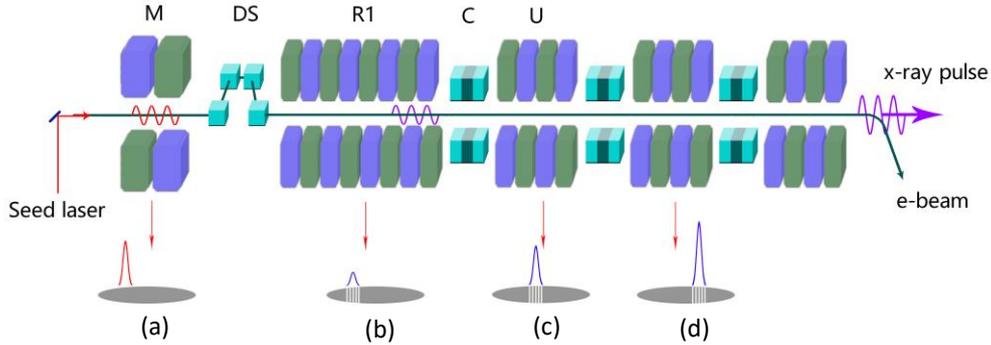

FIG. 1. Schematic layout of the HB-HGHG scheme. (a) A high power seed laser is used to introduce energy modulation into the electron beam in the modulator (M). (b) The energy modulation is converted into density modulation by the dispersion section (DS). The first part of the radiator (R1) is used to generate coherent signal at a target high-harmonic of the seed. (c, d) The following radiator that consists of chicane-undulator (C-U) modules are used to amplify the coherent signal.

Fig. 1 shows the schematic layout to realize the proposed HB-HGHG scheme, which consists of a conventional CHG and an amplifier with a series of chicane-undulator modules. An ultra-short UV seed laser pulse is adopted to interact with a small part of the electron beam in the modulator (M) to generate a sinusoidal energy modulation (Fig. 1 (a)). This energy modulation is then converted into an associated density modulation by the dispersion section (DS) and subsequent coherent emission at a target high harmonic of the seed can be achieved in the following undulator (R1), as shown in Fig. 1 (b). The density modulation of the electron beam can be quantified by the bunching factor, which has a maximal value of unity [15]:

$$b_n = J_n(nDA)\exp(-\frac{n^2 D^2}{2}), \qquad (1)$$

where $J_n$ is the $n$th order first class Bessel function, $D = k_s R_{56} \sigma_\gamma / \gamma$ is the dimensionless parameter related to the dispersive strength of the DS, $k_s$ is the wavenumber of the seed laser, $R_{56}$ is the momentum compaction generated by the DS, $\gamma$ is the energy of the electron beam, $\sigma_\gamma$ is the initial uncorrelated energy spread, $A = \Delta\gamma / \sigma_\gamma$ and $\Delta\gamma$ is the energy modulation amplitude induced by the seed laser. In order to optimize the $n$th harmonic bunching factor, the product of $n$ and $D$ in the exponential term of Eq. (1) should be small enough to limit the energy spread effect, while the product of $n$, $D$ and $A$ should be sufficient large to give a considerable value of $J_n$. Here we assume $nD = 1$ and make $J_n$ reach its first and absolute maximum by using $nDA \approx n+1$, which means that $\Delta\gamma$ should be approximately $n$ times larger than $\sigma_\gamma$. However, when the laser induced energy spread is too large, e.g. larger than the Pierce parameter $\rho_r$ [32] in the radiator, there will be no exponential growth of the FEL and results in a very low output power. Therefore the need to limit the growth of the energy spread prevents the possibility of reaching short wavelength in a single stage HGHG.

The FEL gain process in the radiator of HGHG can be divided in to three stages: the CHG with quadratic growth, the exponential growth and saturation. In the first two gain lengths of the radiator, the FEL works in the CHG regime, where the harmonic field grows linearly with the distance traversed in the radiator $z$, and the peak power grows as $z^2$. For a longitudinal uniform distributed electron beam with current of $I$ and rms transverse beam size of $\sigma$, the output power of a CHG with radiator length of $l_r$ can be simplified as [33]

$$P_{coh} = \frac{(Z_0 K [JJ]_1 l_r I b_n)^2}{32 \pi \sigma^2 \gamma^2}, \qquad (2)$$

where $Z_0 = 377\Omega$ is the vacuum impedance, $K$ is the dimensionless undulator parameter, $[JJ]_1$ is the polarization modification factor for a linearly polarized planar undulator. One can find from Eq. (2) that the output power of CHG is strongly coupled with the beam current, bunching factor and transverse beam size but nearly independent of the initial beam energy spread. Thus we can use CHG to generate coherent signal with relatively low peak power but at ultra-high harmonics of the seed in a short radiator. This coherent signal is then shifted forward to a fresh part of the electron beam by a small chicane inserted between undulator sections. The microbunchings formed in the previous undulator section are smeared out by the chicane, preventing the growth of noisy spikes in the following undulators. The coherent signal reseeds the fresh bunch in the following undulator section, leading to an amplification of the radiation pulse until saturation. This procedure is repeated in the following chicane-undulator modules, as shown in Fig. 1 (c) and (d). The final output power of the radiation pulse can be much higher than the saturation power of a conventional seeded FEL at the same wavelength.

### III. REQUIREMENTS FOR THE SEED LASER AND DELAY CHICANES

The main purpose of the proposed HB-HGHG is to produce nearly transform limited x-ray pulses with very high peak power. However, harmonic multiplication process also amplifies the initial electron shot noise, which may overwhelm the external seeding source. In order to get sufficient contrast in the HB-HGHG, the initial seed laser power $P_{seed}$ should be much higher than the equivalent shot noise power in the modulator, which can be estimated by [34, 35]:

$$P_{noise} = \frac{3^{3/4} 4\pi \rho_m^2 P_{beam}}{N_{\lambda s} \sqrt{\pi l_m / L_{gm}}}, \qquad (3)$$

where $P_{beam}$ is the electron beam power, $N_{\lambda s}$ is the number of electrons per seed laser wavelength, $l_m$ is the modulator length, $\rho_m$ and $L_{gm}$ are the Pierce parameter and FEL gain length in the modulator,

respectively. In order to get a clean output spectrum, the following requirement should be satisfied [36]:

$$P_{seed} / P_{noise} \gg n^2 . \tag{4}$$

This requirement can be easily satisfied for a soft x-ray FEL since the harmonic number is not very large and the high power of the seed laser is not a constrain for the proposed scheme.

The degradation of the output coherence of the HB-HGHG may also arise from the signal to noise ratio of the radiation pulse from CHG, which acts as a start-up seed for the following chicane-undulator stages. The shot noise power generated in the R1 can be estimated by [37]

$$P_{out} = \frac{1}{9} P_{st} \exp(l_r / L_{gr}) . \tag{5}$$

where $P_{st}$ is the start-up shot noise of SASE and $L_{gr}$ is the gain length of the FEL in R1. The power of the coherent signal from CHG is determined by Eq. (2). One can find that the CHG power is proportional to the square of $b_n$, which means that a larger $\Delta\gamma$ (higher seed laser power) will help one to enhance the CHG output and suppress the shot noise in R1. The requirement for the signal to noise ratio of the radiation from R1 is

$$P_{coh} / P_{st} \gg 1 . \tag{6}$$

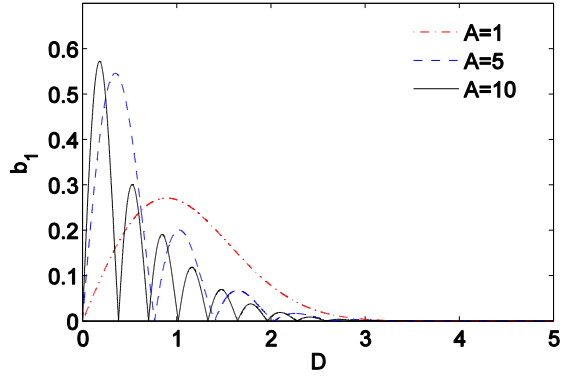

Fig.2. The fundamental bunching factor as a function of the delay chicane strength for different energy modulation amplitudes.

Besides the shot noise, the final spectral noise may also come from the disturbed part of the electron beam which has already lased in the previous undulator sections and contains considerable coherent microbunchings at the radiation wavelength scale. One of the important functions of the delay chicane is to smear out these microbunchings. The bunching factor at the radiation wavelength as a function of the delay chicane strength can also be calculated by Eq. (1) with $k_s$ replaced by $k_r$, which is the wavenumber of the radiation. The calculation results are illustrated in Fig. 2. For a larger energy modulation amplitude, the bunching factor will oscillate as the dispersion strength is increased. However, for a relative large $D$, all the bunching factors will drop quickly, e.g. for $D=4$, all the bunching factors for the three cases will drop below $5 \times 10^{-5}$, which is at the shot noise level.

### IV. OPTIMIZATION AND SIMULATIONS FOR HB-HGHG

In order to illustrate the possible performance with realistic parameters and show the optimization method of HB-HGHG, we carried out start-to-end simulations with the nominal parameters of the SXFEL user facility, as shown in Table. 1. The SXFEL is initially designed for generating coherent soft x-ray in the "water window" (the wavelength of electromagnetic spectrum range from 2.2nm~ 4.4nm) with a 1.6

GeV electron beam based on a two-stage cascaded HGHG setup. The total harmonic up-conversion number for the two stages is 12×5. By using the HB-HGHG scheme, this task could be performed through a single stage configuration.

Table. 1 Main parameters for SXFEL user facility based on HB-HGHG

| *Electron beam* | |
|---|---|
| Energy | 1.6 GeV |
| Bunch charge | 500 pC |
| Peak current | 800 A |
| Energy spread | 160 keV |
| Emittance | 0.6 |
| Full bunch length | 600 fs |
| *Seeding laser* | |
| Wavelength | 265 nm |
| Peak power | 4 GW |
| Pulse length (FWHM) | 30 fs |
| Rayleigh length | 2.96 m |
| *Modulator* | |
| Period | 8 cm |
| Period number | 20 |
| *Dispersion section* | |
| $R_{56}$ | 8 μm |
| *R1* | |
| Period | 2.35 cm |
| Radiation wavelength | 4.4 nm |
| Radiator length | 3.6 m |
| *Chicane-undulator modules* | |
| $R_{56}$ of the chicane | 36 μm |
| Period of the undulator | 2.35 cm |
| Length of one undulator section | 4.3 m |

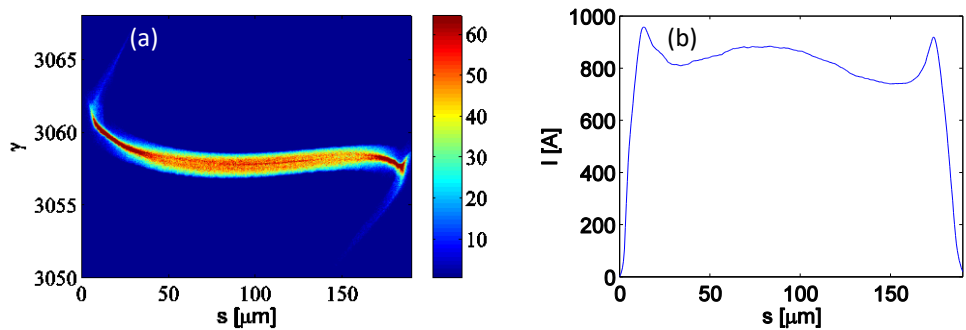

Fig.3. Simulation results for the longitudinal phase (a) and current profile (b) of the electron beam.

The linac of the SXFEL consists of a photoinjector, a laser heater system, an X-band linearizer, S-band and C-band accelerating structures and two magnetic bunch compressors. With the parameters listed in Table. 1, start-to-end tracking of the electron beam, including all components of SXFEL, has been carried out. The simulation of the electron beam dynamics in the photoinjector was performed by ASTRA [38] with space-charge effects taken into account. The electron beam is accelerated to about 130 MeV at the end of the injector. The beam peak current is about 30 A. ELEGANT [39] was then used for the simulation in the remainder of the linac. The uncorrelated energy spread of the electron beam was increased from about 1 keV to about 5 keV by the laser heater system to suppress the microbunching instability in the linac. An X-band linearizer is employed before the first bunch compressor to compensate the second order nonlinear components in the longitudinal phase space to avoid the undesired growth of the transverse emittance and energy spread. Simulation results for the longitudinal phase space and the current profile of the electron beam at the exit of the linac are shown in Fig. 3, where one can find a peak current of over 800 A is achieved and the a constant profile is maintained in the approximately 300 fs wide region. This part of electron beam is used for the FEL amplification in HB-HGHG.

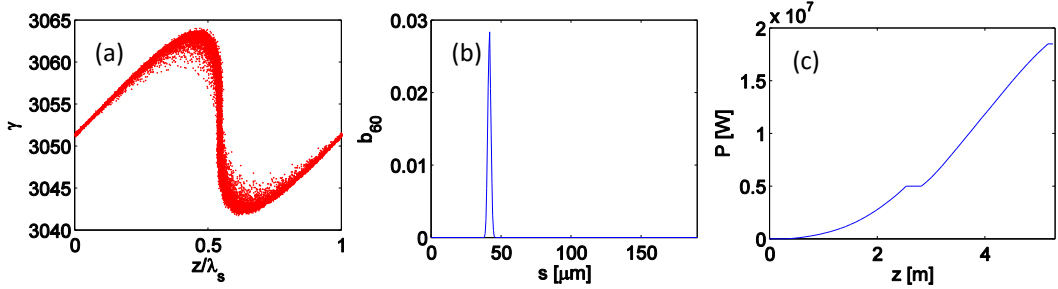

Fig.4. Simulation results for the CHG: (a) longitudinal phase space in one seed laser wavelength region; (b) the 60$^{th}$ harmonic bunching factor distribution along the electron bunch; (c) the gain curve of the FEL peak power in R1.

The modulation and FEL gain processes were simulated with GENESIS [40] based on the output of ELEGANT. An UV seed laser at 265 nm with pulse duration of about 30 fs (FWHM) is adopted to interact with the tail part of the electron beam in a modulator with period length of 8 cm and period number of 20. By using Eq. (3) and the electron beam parameters shown in Table. 1, the shot noise power in the modulator is calculated to be about 100 W. For the 60$^{th}$ harmonic generation, the requirement for the seed laser power is $P_{seed} \gg 0.36 MW$ according to Eq. (4). Here we chose the seed laser peak power of about 4 GW. The laser induced energy modulation amplitude is about 6.1 MeV, which is about 38 times of the uncorrelated beam energy spread. A dispersion section with R$_{56}$ of about 8 μm is employed to convert density modulation in to density modulation. The simulation results for the longitudinal phase space and corresponding bunching factor distribution along the electron bunch are shown in Fig. 4 (a) and (b). The bunching factor at 60$^{th}$ harmonic of the seed is over 2.5%, which is responsible for the initial steep quadratic growth of the radiation power in R1, as shown in Fig. 4 (c). The output peak power CHG at 4.4 nm is about 18 MW. From the simulations, we also found that the power of the SASE at the exit of R1 is about 8000 W. Thus the signal to noise ratio of the radiation pulse from R1 is about $P_{coh}/P_{st} \approx 2250$, which is sufficient large to suppress the shot noise effect on the final output coherence.

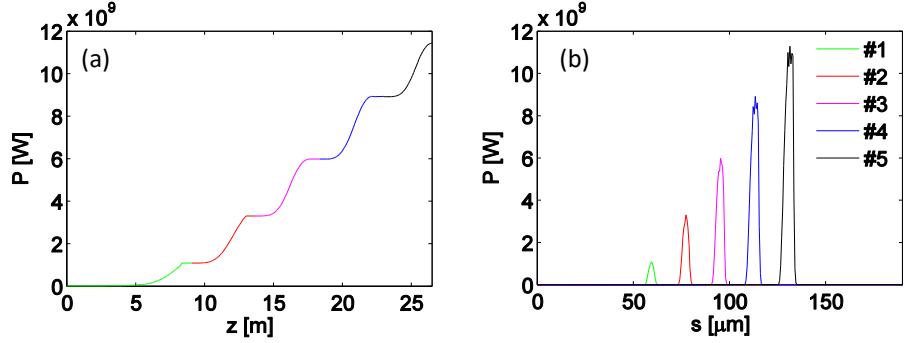

Fig. 5. The FEL gain curve (a) and structure evolutions of the radiation pulse (b) in the chicane-undulator modules.

The CHG signal is then shifted forward to the fresh part of the electron beam and sent into the downstream chicane-undulator modules to get further amplifications. The $R_{56}$s of the delay chicanes are set to be 36 μm, which can introduce relative time delays of about 60 fs and dimensionless dispersion strength $D \approx 5.14$. The dispersion strengths of the delay chicanes are large enough to smear out the microbunchings formed in the previous undulator sections. 5 chicane-undulator modules were used in the simulation. As the initial seed from CHG is quite weak, a relative long undulator section with length of about 8 m is adopted in the first chicane-undulator module to fully amplify the CHG signal. The lengths of the rest chicane-undulator modules are all set to be about 4.5 m, which is enough to place a 3.8 m long undulator and a 0.5 m long delay chicane. The undulator parameters are optimized in each module to enhance the output power. The simulation results for the FEL gain curve and the radiation pulse evolutions in the chicane-undulator modules are summarized in Fig. 5. The peak power of the radiation pulse from the first undulator section is about 1.1 GW, which is already comparable to the saturation power of a two-stage cascaded HGHG FEL at the same wavelength. This intense radiation pulse can be further amplified by the following chicane-udulator modules via the superradiant process. The output peak power can be enhanced well beyond saturation. As shown in Fig. 5, the output peak power is increased to about 11.5 GW after 5 modules.

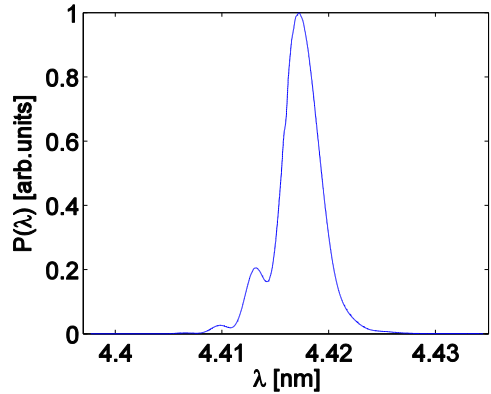

Fig. 6. The output spectrum of the HB-HGHG.

It is anticipate that seed FELs can generate nearly transform-limited radiation pulse, which usually requires a uniform electron beam with nearly constant energy and current distributions. However, the energy profile of the electron beam usually has an initial energy curvature due to the radio frequency curvature, wakefield effects and microbunching instabilities in the linac. Generally, this energy curvature will impact the FEL gain process and result in a broader output spectral bandwidth [41, 42]. When compared with the conventional cascaded HGHG scheme, the proposed scheme only needs one DS with much smaller dispersion strength due to the relatively large energy modulation amplitude. Thus means that the effects of the imperfect energy profile on the output coherence can be significantly reduced. This is considered as another advantage of HB-HGHG for generating ultra-high harmonic radiations. The simulation result for the final output spectrum is shown in Fig. 6. The FWHM bandwidth of the spectrum is about 0.9%, which is about 1.4 times of the Fourier transform-limit.

## V. CONCLUSIONS

In conclusion, the HB-HGHG scheme has been proposed in this paper for enhancing the harmonic up-conversion efficiency and the output intensity of a seeded FEL. The proposed scheme is simple and easy to be implemented at seeded FEL facilities. In addition to a standard HGHG scheme, it only needs small delay chicanes between undulator sections. The feasibility of operating HB-HGHG for the generation of high brightness and coherent radiation pulses in the water window has been demonstrated by numerical simulations. The simulation results show that the proposed scheme improves the output peak power by about one order of magnitude with respect to a conventional HGHG. The output peak power can be further increased by using more chicane-undulator modules and a longer electron beam. The short-wavelength cover range can also be extended by using a seed laser pulse with higher peak power. This kind of light source would allow one to perform many challenging experiments which require narrow bandwidth soft x-ray radiation pulses.

## ACKNOWLEDGEMENTS

The authors would like to thank Dazhang Huang, Zhen Wang, Zheng Qi and Kaiqing Zhang for helpful discussions and useful comments. This work is supported by the National Natural Science Foundation of China (11475250) and Youth Innovation Promotion Association CAS.